\documentclass[usenatbib,useAMS]{mn2e}
\usepackage{natbib}
\usepackage{amsmath}
\usepackage{amsfonts}
\usepackage{multirow}
\usepackage{array}
\usepackage{url} 
\usepackage{epsfig}
\usepackage{verbatim} 
\usepackage[usenames]{color}
\allowdisplaybreaks[1]

\definecolor{purple}{RGB}{160,32,240}

\newcommand{\peter}[1]{}

\newcommand{\Msol}{{\mathrm M}_{\bigodot}}

\newcommand{\K}{{\rm K}}

\newcommand{\kB}{k_{\rm B}}
\newcommand{\km}{{\rm km}}
\newcommand{\cm}{{\rm cm}}
\newcommand{\s}{{\rm s}}
\newcommand{\kms}{{\rm km}\,{\rm s}^{-1}}
\newcommand{\Myr}{{\rm Myr}}

\newcommand{\Mpc}{\rm{Mpc}}

\newcommand{\rvir}{r_\mathrm{vir}}
\newcommand{\Tvir}{T_\mathrm{vir}}

\newcommand{\Hmol}{\mathrm{H_2}}

\newcommand{\beq}{\begin{equation}}
\newcommand{\eeq}{\end{equation}}

\newcommand{\apj}{ApJ}
\newcommand{\apjl}{ApJL}
\newcommand{\apjs}{ApJS}

\newcommand{\aapr}{A$\&$A}
\newcommand{\jcap}{J. Cosmol. Astropart. Phys.}

\newcommand{\araa}{ARA$\&$A}

\newcommand{\mnras}{MNRAS}

\newcommand{\prd}{Phys. Rev. D}
\newcommand{\aj}{AJ}
\newcommand{\nat}{Nature}
\newcommand{\nar}{New Astron. Rev.}
\newcommand{\physrep}{Physics Reports}

  \def\spose#1{\hbox to 0pt{#1\hss}}
  \def\ltsim{\mathrel{\spose{\lower.5ex\hbox{$\mathchar"218$}}
  \raise.4ex\hbox{$\mathchar"13C$}}}

\title[Early massive BHs from large streaming velocities]
{The formation of massive black holes in $z\sim 30$ dark matter haloes with large baryonic streaming velocities}

\author[T.L.Tanaka and M. Li]
{Takamitsu L. Tanaka$^{1}$\thanks{E-mail: taka@mpa-garching.mpg.de}, Miao Li$^{2}$\\
$^{1}$Max Planck Institute for Astrophysics, Karl-Schwarzschild-Str. 1, D-85741 Garching, Germany\\
$^{2}$Department of Astronomy, Columbia University, 550 W. 120th Street, New York, NY 10027, USA}
\begin{document}

\maketitle
\label{firstpage}

\begin{abstract}
The origins of the $\sim 10^9\,\Msol$ quasar supermassive black holes (BHs) at redshifts $z>6$ remain a theoretical puzzle.
One possibility is that they grew from $\sim 10^5\,\Msol$ BHs
formed in the `direct collapse' of pristine, atomic-cooling (temperatures $\ga 8000 \,\K$; PAC) gas
that did not fragment to form ordinary stars due to a lack of molecular hydrogen and metals.
We propose that baryonic streaming---the relic relative motion of gas with respect to dark matter from
cosmological recombination---provides a natural mechanism for establishing the
conditions necessary for direct collapse.
This effect delays the formation of the first stars by inhibiting
the infall of gas into dark matter haloes;
streaming velocities more than twice the root-mean-square value could forestall
star formation until halo virial temperatures $\ga 8000 \,\K$.
The resulting PAC gas can proceed to form massive BHs by any of the mechanisms
proposed in the literature to induce direct collapse in the absence of a ultraviolet background.
This scenario produces haloes containing PAC gas at a characteristic redshift $z\sim 30$.
It can explain the abundance of the most luminous quasars at $z\approx 6$,
regardless of whether direct collapse occurs in nearly all or less than 1 per cent of PAC haloes.
\end{abstract}


\section{Introduction}
\label{sec:intro}

Observations of quasars at redshifts $z\approx 6\,$--$\,7$
reveal that supermassive black holes (SMBHs)
with masses $M_{\rm SMBH}>10^9\,\Msol$
had already formed less than 1 Gyr after the Big Bang \citep{Willott+03,Fan06, Mortlock+11}.
The formation mechanism of these objects remains
an open theoretical question (see reviews by \citealt{Volonteri10} and \citealt{Haiman13}).

One hypothesis is that these objects grew from the remnants of the
first (Population III or `PopIII') stars (e.g. \citealt{MadauRees01}, \citealt{HaimanLoeb01},
\citealt{Shapiro05}, \citealt{Li+07}, \citealt{Pelupessy+07}),
which formed as gas collapsed through molecular hydrogen ($\Hmol$) cooling
inside dark matter (DM) haloes with virial temperatures $\Tvir\sim 1000\,\K$ 
at $z\sim 20\,$--$\,40$
(see the review by \citealt{BrommYoshida11}).
PopIII stars may have masses as large as $300\,\Msol$ \citep{Heger+03,OmukaiPalla03,Ohkubo+09}
and as small as $\sim 10\,\Msol$ \citep{Turk+09,Stacy+10,Clark+11,Greif+11b,Hosokawa+11},
leaving BHs with  $\sim 40$ per cent of the stellar mass \citep{ZhangW+08}.

The minimum time-averaged mass accretion rate
for PopIII remnants to grow into the observed $z>6$ SMBHs
can be written as a fraction of the canonical Eddington limit,
by comparing the number of required $e$-foldings gained via accretion
to the amount of $e$-folding times available:
\begin{align}
f_{\rm Edd}\ga& \ln \left(\frac{M_{\rm SMBH}}{X_{\rm merge}~M_{\rm seed}}\right)\Big/
\left[\frac{t_{\rm avail}}{(\eta/0.07)\,t_{\rm Edd}}\right]\nonumber\\
\approx&
\left[0.676+0.045\ln\left(\frac{M_{\rm SMBH}}{3\times 10^9\,\Msol} \frac{30\,\Msol}{M_{\rm seed}}\frac{30}{X_{\rm merge}}\right)
\right]\nonumber\\
&\qquad \times\left(\frac{\eta}{0.07}\right)\left(\frac{t_{\rm avail}}{700\,\Myr}\right)^{-1}.
\label{eq:fEddPopIII}
\end{align}
Here, $M_{\rm seed}$ is the mass of the `seed' BH;
$X_{\rm merge}\sim 10-10^3$ \citep{TH09, TPH12}
is the growth by BH mergers via hierarchical structure formation;
$(\eta/0.07)\,t_{\rm Edd}=31.5\,\Myr$
is the $e$-folding timescale for a radiative efficiency $\eta\equiv L/(\dot{M}_{\rm BH}c^2)$
for luminosity $L$ and accretion rate $\dot{M}_{\rm BH}$
scaled to the value derived by \cite{MerloniHeinz08} and \cite{Shankar+09a} 
see also \cite{Shapiro05};
and $t_{\rm avail}\approx 700\,\Myr$ is the available time from the formation of the earliest seeds
at $z\ga 40$ until $z\approx7$.
It is unclear whether such a large accretion rate can be sustained
for such a long time.
Interestingly, empirical measurements of the quasar duty cycle
are indeed as high as $\ga 0.5$ \citep{Shankar+09a,Willott+10b};
the fact that the most massive haloes at these redshifts grow
far more rapidly than in the local Universe \citep{Angulo+12}
could explain such prolific accretion activity.
However, negative radiative feedback could reduce
the accretion rate to a minuscule fraction of the required value,
especially at early stages when the gravitational potential
of the host halo is shallow \citep{Alvarez+09,Milos+09}.

An alternate path to SMBH formation is the `direct collapse'
of a gas cloud of nearly primordial composition
and temperature $T\ga 10^4\,\K$
\citep{OH02, BrommLoeb03,Koushiappas+04,LodatoNatarajan06, SpaansSilk06, Dijkstra+08, ReganHaehnelt09a, ReganHaehnelt09b, Shang+10, Agarwal+12, Latif+13}.
The paramount prerequisite for this scenario is that the fractions of
$\Hmol$, metals and dust---strong coolants that trigger
fragmentation into stars---be kept minimal.
If this condition is satisfied, then compressional heating
can balance cooling via atomic transitions.
The cloud collapses nearly isothermally at $T\ga 8000\,\K$
to form a $\sim 10^{5}\,\Msol$ star or star-like massive envelope \citep{Begelman+06, Hosokawa+13,Schleicher+13}
that ultimately leaves behind a massive BH with a similar mass \citep{ShibataShapiro02, LN07, Latif+13b}.

To host such hot gas, the potential of the DM halo
must be deep---i.e. its virial temperature
must be $\Tvir\ga 8000\,\K$.
However, as stated above,  $\Hmol$ cooling usually triggers PopIII formation at a typical
virial temperature value $\Tvir\sim 1000\,\K$.
The supernovae of these stars would distribute metals and dust,
particularly if some PopIII stars are $\sim 100\,\Msol$ \citep{Greif+10};
a single pair-instability supernovae may be sufficient
to trigger the transition to Population II stars \citep{Wise+12}.
Therefore, in order for direct collapse to occur,
the host halo must experience minimal (massive) star formation from when
$\Tvir\sim 1000\,\K$ until $\Tvir\ga 8000\,\K$,
a gap corresponding to a factor of $\ga 20$ growth in halo mass.

A strong UV background,
such as from nearby star-forming galaxies or quasars,
could suppress the $\Hmol$ fraction through photodissociation,
prevent PopIII formation as the halo accumulates mass,
and lead to direct collapse once the halo reaches $\Tvir\sim 10^4\,\K$
\citep{OH02, BrommLoeb03, Dijkstra+08, Shang+10}.
The time-averaged accretion rate required to grow into the most massive $z>6$
SMBHs is lower for direct-collapse remnants than for PopIII seeds,
but this is still a significant fraction of the Eddington limit:
\begin{align}
f_{\rm Edd}\ga
&
\left[0.580+0.063\ln\left(\frac{M_{\rm SMBH}}{3\times 10^9\,\Msol} \frac{10^5\,\Msol}{M_{\rm seed}}\frac{3}{X_{\rm merge}}\right)
\right]\nonumber\\
&\qquad \times\left(\frac{\eta}{0.07}\right)\left(\frac{t_{\rm avail}}{500\,\Myr}\right)^{-1}.
\label{eq:fEddUV}
\end{align}
Here, we have scaled the beginning redshift to $15$
\citep[approximately the earliest time at which stars or quasars can build up
a $\Hmol$-dissociating UV background; e.g.][]{Agarwal+12}
and reduced the factor $X_{\rm merge}$ to account for the rarity of massive BH seeds \citep{TH09}.
Comparing equations \ref{eq:fEddPopIII} and \ref{eq:fEddUV}, we see
that while massive BH seeds formed from UV-aided direct collapse require
a lower mean accretion rate than PopIII remnants,
this rate still must be a significant fraction of Eddington
for most of the age of the Universe at $z=7$.

Haloes with $\Tvir\ga 8000\,\K$ can contain two different
phases of gas \citep{BirnboimDekel03, Keres+05, DekelBirnboim06,Dekel+09}. 
Diffuse gas that is just below $8000\,\K$ will be unable
to cool efficiently, as atomic cooling is inefficient
and such gas will have low $\Hmol$ densities \citep[e.g.][]{OH02}.
Gas that is denser will have a higher $\Hmol$
fraction and cool slightly faster, increase its $\Hmol$ fraction, etc. in a runaway fashion.
Thus, atomic-cooling haloes can contain
dense cold filaments embedded in diffuse, hot gas.
The filaments can sink to the halo center at supersonic velocities,
and may play a central role in SMBH growth at high redshifts
by delivering large supplies of dense gas \citep{Greif+08, DiMatteo+12}.
Supersonic turbulence may enhance the formation and velocities of cold filaments
\citep{WiseAbel07, Greif+08, Wise+08, Prieto+13}.

Several studies have proposed ways in which direct collapse may
occur in the absence of a UV background.
\textit{Collisional} dissociation can keep $\Hmol$ fractions low if
the gas is hot ($T\ga 8000\,\K$) and dense ($n\ga 10^3\,\K$).
Additionally, under such conditions the $\Hmol$ roto-vibrational levels saturate
to local thermodynamic equilibrium, reducing
the net cooling rate per molecule \citep{IO12}.
The  gas could stay at a temperature of $\sim 8000\,\K$
without forming $\Hmol$ if the neutral hydrogen column density
is large enough to trap Lyman$\alpha$ cooling radiation \citep{SpaansSilk06,SSG10, Latif+11}.
Gravitational instabilities could transport angular momentum efficiently
and lead to direct collapse \citep{BegelmanShlosman09}.
Gas can be more efficiently delivered to the deepest
part of the potential if its angular momentum is low
\citep{EisensteinLoeb95, Koushiappas+04, LodatoNatarajan06}.
Recently, \cite{IO12} proposed that direct collapse could be triggered by cold-accretion filaments shocking at the center of the halo.
In each of these scenarios, the atomic-cooling, $\Hmol$-free
cloud can collapse monolithically, much as in
the UV background-aided picture of direct collapse.
There are two significant theoretical uncertainties.
First, prior PopIII formation and metal enrichment may facilitate cooling
and fragmentation, preventing direct collapse.
Again, the halo must grow by a factor $\ga 20$ in mass
without becoming significantly metal enriched.
Second, these scenarios have not been thoroughly tested by
detailed hydrodynamical simulations.

In this work, we propose that baryonic streaming motions (BSMs)---the
velocity of baryons relative to DM at cosmological recombination \citep{TseliakHirata10}---provide
a natural mechanism for forestalling star formation and keeping the gas pristine
until direct collapse can occur in any of the scenarios listed above.
BSMs impede the infall of gas into early DM haloes,
thus delaying PopIII formation until the velocities decay (as $\propto 1+z$)
and the haloes have deeper potentials \citep{Stacy+11,Greif+11a,Naoz+13}.
For typical values of the streaming velocity---root-mean-square (rms) speed
$\sigma_{\rm BSM}^{\rm (rec)} \approx 30\,\km\,\s^{-1}$ at recombination ($z\approx1000$)---BSMs
delay the formation of PopIII stars until the host halo mass has tripled (\citealt{Greif+11a}; see also \citealt{Stacy+11}),
compared  to a theoretical situation where the velocity is zero.
This delay has only a small impact on the globally averaged histories of
reionization and SMBH formation\footnote{
BSMs may leave detectable imprints
in the power spectra of galaxies, quasars and the 21 cm signature
\citep{Dalal+10, Maio+11, Tseliak+11, McQuinnOleary12, Visbal+12}.}
\citep{TLH13}\footnote{
Note that whereas \cite{TLH13} discussed the (weak) negative
effect of BSMs on the formation of SMBHs from PopIII seeds,
this paper discusses how BSMs could play a positive role
by enabling the early formation of direct-collapse BHs.
}.

We will show that in very rare patches where the streaming
velocities are more than twice the rms value, the most massive haloes
at $z\sim 30$ could reach $\Tvir\ga 8000\,\K$
before forming PopIII stars.
Because gas falling into them would be pristine and have sufficiently
large temperatures to be in the atomic-cooling regime,
these exceptionally rare haloes would be natural sites
where direct collapse could occur very early, without a UV background.
We term these sites pristine atomic-cooling (PAC) haloes.
We show that early formation of massive BHs via direct collapse in PAC haloes
can explain the abundance of the most massive quasar BHs at $z\approx 6$--$7$,
regardless of whether direct collapse occurs generically
or very rarely (e.g. in less than one percent of cases) in $z\sim 30$ PAC haloes.

This paper is organized as follows.
In \S\ref{sec:masses}, we present simple analytic arguments
to show that at suitably high streaming velocity values,
DM haloes can reach the atomic-cooling threshold
before forming PopIII stars.
We estimate in \S\ref{sec:numbers}
the comoving number density of massive BHs
formed in this way.
We discuss several theoretical considerations and offer
concluding remarks in \S\ref{sec:disc}.

Throughout this work, $c$, $G$, $\kB$ and $m_{\rm p}$
denote the speed of light, the gravitational constant,
the Boltzmann constant and the proton mass, respectively.
Cosmological parameters for a $\Lambda$ cold dark matter ($\Lambda$CDM)
universe are denoted in the usual way:
$h$, $\Omega_0$, $\Omega_\Lambda$, $\Omega_{\rm b}$, $n_{\rm s}$,
$\sigma_8$.

\section{Halo mass threshold for gas infall and PopIII formation}
\label{sec:masses}
In the absence of baryonic streaming, $\Hmol$ forms
in gas accumulating inside haloes with 
$T_{\rm vir}\sim 400\,$--$\,1000\,\K$ \citep{Haiman+96, Tegmark+97};
the gas then collapses to form stars.
This threshold can be expressed in terms
of the halo circular velocity or halo mass:
\begin{align}
v_{\rm circ}&\equiv \sqrt{\frac{GM}{\rvir}}=3.7
\left(\frac{T_{\rm vir}}{1000\,\K}\right)^{1/2}
\left(\frac{\mu}{1.2}\right)^{-1/2}
\km\,\s^{-1};\\
 M&=2.6\times 10^5 
 \left(\frac{T_{\rm vir}}{1000\,\K}\right)^{3/2}\left(\frac{1+z}{26}\right)^{-3/2}
\nonumber\\
&\qquad\times \left(\frac{h}{0.7}\right)^{-1} \left(\frac{\Omega_0}{0.27}\right)^{-1/2}\left(\frac{\mu}{1.2}\right)^{-3/2}\,\Msol
\label{eq:Mvir},
\end{align}
where $\rvir$ is the halo virial radius
and $\mu$ is the mean molecular weight \citep[e.g.][]{BarkanaLoeb01}.

BSMs delay gas infall, and increase the typical halo mass at which PopIII stars form.
By analyzing the results of hydrodynamical
simulation of PopIII formation that included this effect,
\cite{Fialkov+12} fit the new circular velocity
threshold $v_{\rm cool}$ to the following analytic form:
\begin{equation}
v_{\rm cool}=\sqrt{v_{0}^2 + \left[\alpha \, v_{\rm BSM}(z)\right]^2}.
\label{eq:vcool}
\end{equation}
They arrived at parameter values of $(v_0,\,\alpha)=(3.64\, \kms, 3.18)$
and $(3.79\,\kms, 4.71)$ for the results of \cite{Stacy+11}
and \cite{Greif+11a}, respectively.

Naively, one might expect BSMs to negatively impact
gas infall when $v_{\rm BSM}\sim v_{\rm circ}\approx 3.7\,\kms$,
i.e. that $\alpha\sim 1$.
Instead, the simulations show that $\alpha \sim 4$,
i.e. that streaming velocities that are only a fraction of the circular velocity
is sufficient to delay PopIII formation.
Plausibly, this is because the relevant velocity
value is not the halo's circular velocity when stars finally form,
but rather the value related to the infall
of the gas at earlier times (when streaming motions are larger and
the halo potential is shallower) and at a radius larger than
the virial radius.
\cite{Stacy+11} suggested that the relevant velocity is the sound speed
of the gas in the intergalactic medium when the gas first begins
to fall into the halo potential, i.e. that BSMs raise the Jeans mass scale.
\cite{Naoz+13} showed that the filtering mass scale
\citep{Gnedin00, NaozBarkana07}, which takes into account
the thermal history of the gas, can more accurately explain
the characteristic mass scale.

Additionally, the enhancement of two heating mechanisms
could contribute to the delay in PopIII formation.
\cite{Yoshida+06} found in their simulations that haloes with large mass
accretion rates did not form PopIII stars right away,
due to greater dynamical heating by the accreting matter.
This heating rate is
\beq
(\gamma-1)\frac{{\rm d}e}{{\rm d}t}\sim \frac{{\rm d}\kB T_{\rm vir}}{{\rm d}t}\approx \frac{\mu m_{\rm p} G}{3\rvir}\dot{M}_{\rm halo}
~~({\rm dyn.~heating})
\eeq
where $\gamma$ is the adiabatic index and
$e\equiv (\gamma-1)^{-1}\kB T/(\mu m_{\rm p})$ is the internal energy of the gas per unit baryonic mass.
Because on average the halo mass accretion rate is roughly
proportional to the halo mass \citep[e.g.]{Wechsler+02, Fakhouri+10}\footnote{Equation 1
in \cite{Fakhouri+10} suggests $\dot{M}_{\rm halo}\sim 0.2 M_{\rm halo}~{\rm d}z/{\rm d}t$
within a factor of two for $10^5\,\Msol<M<10^9\,\Msol$ and $\Omega_0(1+z)^3\gg \Omega_{\Lambda}$.},
dynamical heating would be, on average, stronger in haloes where BSM delays gas accumulation.
Similarly, the compressional heating rate of a collapsing gas cloud
scales with the free-fall timescale $t_{\rm ff}=\sqrt{3\upi/(32G\bar{\rho})}$ \citep[e.g.][]{Omukai+08}:
\beq
\frac{{\rm d}e}{{\rm d}t}=p\frac{{\rm d}\ln \rho_{\rm gas}}{{\rm d}t}
\sim \frac{\kB T}{\mu m_{\rm p}}\frac{1}{t_{\rm ff}}\propto \bar{\rho}^{-1/2}
~~({\rm comp.~heating})
\eeq
where $p$ is the gas pressure and $\bar{\rho}$
is the combined mean density of gas and DM.
In haloes where gas infall is delayed, the DM-to-gas
ratio is initially greater than in haloes where this ratio
is comparable to the cosmological average,
and thus the compressional heating rate is higher (the free-fall time is shorter)
at similar gas densities.

In what follows, we adopt the fitting formula of \citeauthor{Fialkov+12}
(\citeyear{Fialkov+12}; equation \ref{eq:vcool}), 
accepting that it may be inaccurate by a factor of order unity
compared to more explicit formulations of the characteristic mass
for gas infall \citep{Naoz+13}.
We take $v_0=3.7\, \kms$  and treat $\alpha$ as a free parameter
of order $\sim 4$ as found by \cite{Fialkov+12}.
(Note that in the high-$v_{\rm BSM}$ regime of interest,
the halo mass threshold is much more sensitive to $\alpha$
than to $v_0$.)
The halo mass at which PopIII stars can form is
\begin{align}
M_{\rm cool} \approx
 3.5\times 10^5 
 &\left[1+\left(\frac{2}{3}\frac{\alpha}{4}\frac{v_{\rm BSM}^{\rm (rec)}}{30\,\km\,\s^{-1}}\frac{1+z}{21}\right)^2\right]^{3/2}
 \nonumber\\
&\qquad\times\left(\frac{1+z}{21}\right)^{-3/2}\Msol.
\label{eq:Mcool}
\end{align}
At adequately large velocities, gas infall and PopIII formation are
delayed until the halo potential is deep enough to host
atomic-cooling gas, i.e. $M_{\rm infall}>M(T_{\rm vir}=8000\,\K)\approx 5.8\times 10^6\,\Msol [(1+z)/26]^{-3/2}$ 
(see equation \ref{eq:Mvir})
\footnote{The mass threshold for atomic cooling may be larger by a factor
of order unity (see e.g. \citealt{Fernandez+13}), e.g. if the relevant molecular weight of the gas is $\mu=0.6$ (ionized) as opposed
to $\mu=1.2$ (neutral).}.
Indeed, in their $N$-body simulations \cite{Naoz+13}
find that for $v_{\rm BSM}$ values more than $\sim 2$ times
the rms value, the baryonic content of haloes in this mass range at $z\ga 20$
is lower by a factor of several compared to the case $v_{\rm BSM}=0$.

In the redshift range of interest, we expect the effect of a UV background to be
negligible. Even for optimistic assumptions for the required local Lyman-Werner flux
to induce direct collapse, \cite{Agarwal+12} found that direct collapse does not
occur until $z\approx 16$. 
As we show below (Figure \ref{fig:seeds}),
the PAC halo formation rate peaks at $z\approx 30$,
and falls by several orders of magnitude by $z\approx 16$.
Moreover, our PAC haloes form preferentially in regions of space where prior star formation
is suppressed due to the local streaming velocity 
being faster than the cosmic average; thus, these sites should
anti-correlate with peaks in the UV flux.

\begin{figure}
\epsfig{file=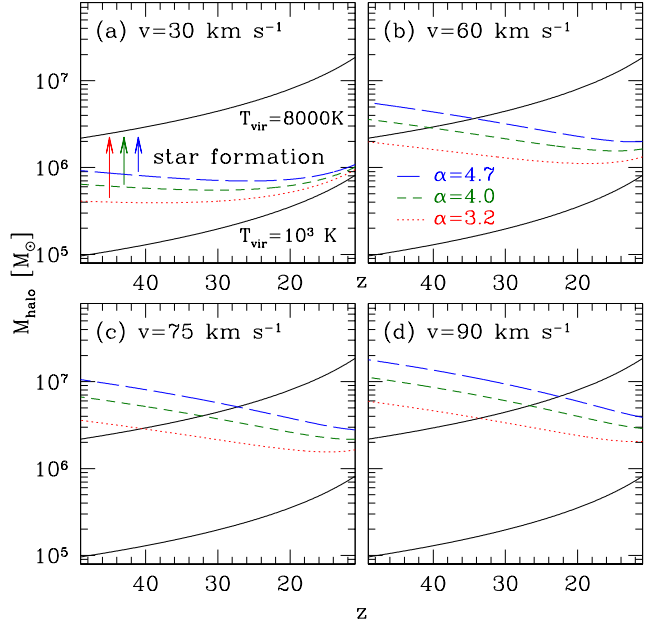,scale=0.44}
\vspace{-0.2in}
\caption{
Characteristic halo mass scales. In each panel, the black curves
show, as a function of $z$, the halo mass corresponding to
$\Tvir=8000\,\K$ (threshold for halo to host atomic-cooling gas)
and to $\Tvir =1000\,\K$ (threshold for PopIII formation in the absence
of streaming velocities).
The colored curves in each panel show the increase threshold
for PopIII formation as estimated by Fialkov et al. (2012; equation 8)
in the presence of various values of the streaming velocity at $z\approx 1000$:
$30\,\kms$ (the rms value), $60\,\kms$, $75\,\kms$ and $90\,\kms$
in panels (a) through (d), in that order.
The colored and styled lines denote different assumptions for the parameter
$\alpha$: 3.2 for red dotted curves, 4.0 for green short-dashed curves
and 4.7 for blue long-dashed curves.
For streaming velocities more than twice the rms value,
the threshold for PopIII formation can exceed the one for atomic cooling.
(A color version of this figure is available online.)
}
\label{fig:masses}
\end{figure}

In Fig. \ref{fig:masses}, we present the halo mass threshold 
for PopIII formation as a function of redshift,
by evaluating equations \eqref{eq:vcool} and \eqref{eq:Mcool}
for streaming velocity magnitudes of
 $v_{\rm BSM}^{\rm (rec)}=30\,\km\,\s^{-1}$ (the rms value),
 $60\,\km\,\s^{-1}$,  $75\,\km\,\s^{-1}$ and  $90\,\km\,\s^{-1}$.
The figure can be read as follows.
The solid curves in each panel show the halo masses corresponding
to virial temperatures of $1000\, \K$ and $8000\, \K$, as labeled in
panel (a). The dotted red, short-dashed green and long-dashed blue curves show
the mass threshold for gas to collapse inside haloes, for
assumed values of $\alpha=3.2$, $4.0$ and $4.7$, respectively;
these values were chosen to span the range in $\alpha$
found by \cite{Fialkov+12}, with $\alpha=4.7$ corresponding
to the high-resolution moving-mesh simulations of \cite{Greif+11a}
At any given $z$ and $v_{\rm BSM}^{\rm ( rec)}$,
haloes with masses below the values indicated by the colored curves
will have gas infall delayed by BSMs, and the gas inside them
will not have collapsed to form PopIII stars.
Haloes with masses above the curved curves will form
compact baryonic objects;
those with virial temperatures $T_{\rm vir}<8000\,\K$ will undergo
molecular cooling and form PopIII stars,
whereas those with $T_{\rm vir}>8000\,\K$
would form PAC clouds and possibly serve as the cradles of massive BHs.

\section{Comoving number density of massive BHs formed via baryonic streaming}
\label{sec:numbers}

The potential direct-collapse sites proposed here are expected to be very rare,
combining two unusual characteristics.
First, they must be very precocious, with masses of $\sim 10^{7}\,\Msol$ 
at $z\ga 20$, and be of order $\sim 10$ times more
massive than the typical contemporary haloes forming PopIII stars.
(Note that even streaming velocities close to the rms value reduce by a factor of a few
the gap in halo mass between PopIII formation and atomic cooling.)
Second, these sites must lie in regions of space where the streaming
velocities are $\ga 2$ times the rms value.
We now turn to estimating the comoving number densities of these sites.

\subsection{Semi-analytic estimates}
\label{subsec:semianalytic}
\begin{figure}
\epsfig{file=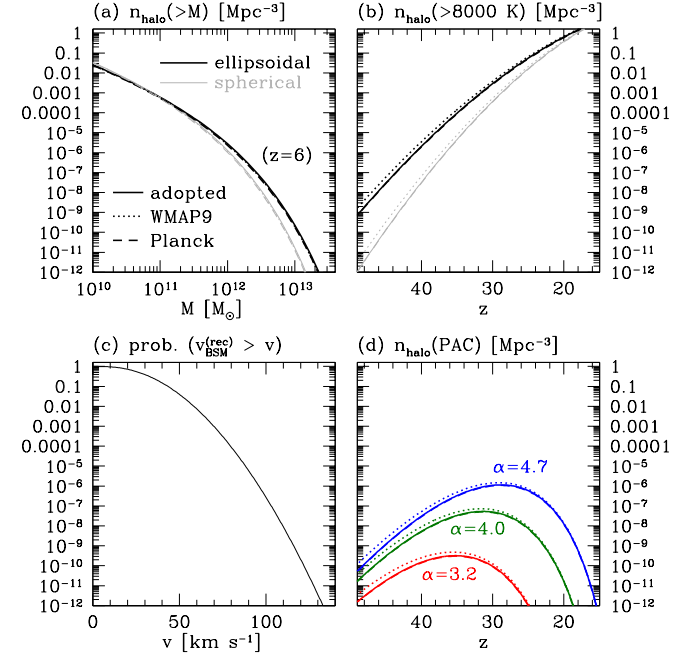,scale=0.44}
\vspace{-0.2in}
\caption{
The rarity of relevant cosmological occurrences.
Panel (a): the comoving number density of the most massive
DM haloes at $z=6$, assuming Sheth-Tormen ellipsoidal collapse (thick black lines)
or Press-Schechter spherical collapse (thin grey lines).
The different line styles correspond to different cosmological parameter sets (Table 1):
solid lines for our parameter choices, dotted lines for \textit{WMAP9} and
dashed lines for \textit{Planck}.
Panel (b): the comoving number density of haloes
with $\Tvir>8000\,\K$ as a function of $z$. Line styles
are as in panel (a).
Panel (c): The likelihood that a given point in space
has a recombination-value of the streaming velocity
above a certain value $v$.
Panel (d): the number density of haloes that
have $\Tvir>8000\,\K$ and also lie in a fast-streaming region
where PopIII formation is prevented.
The colors show different assumed values of the parameter $\alpha$
in the mass threshold estimate in equation 8:
red, green and blue for $\alpha=3.2$, $4.0$ and $4.7$, respectively.
The line styles correspond to different cosmological parameters
as in panels (a) and (b); only the Sheth-Tormen case is shown.
(A color version of this figure is available online.)
}
\label{fig:rarities}
\end{figure}

We quantify in Fig. \ref{fig:rarities} the rarity of several relevant types of objects.
In panel (a), we plot the theoretical comoving number density of the most
massive haloes at $z=6$,
\beq
n_{\rm halo}(>M, z=6)=\int_M^\infty \frac{dn}{dM^\prime} (z=6) ~dM^{\prime}.
\eeq
The solid line shows the results for our adopted cosmological parameters---$h=0.7$,
$\Omega_0=0.3$, $\Omega_\Lambda=0.7$, $\Omega_{\rm b}=0.047$,
$\sigma_8=0.83$ and $n_{\rm s}=0.96$---while the dotted and dashed lines show
the results for parameters according nine-year results of
the \textit{Wilkinson Microwave Anisotropy Probe} (\textit{WMAP}9; \citealt{Hinshaw+13})
and the first data release of the  \textit{Planck} mission \citep{PlanckParameters},
respectively.
The results for the three different sets of parameters effectively overlap,
in this panel as in panels (b) and (d) discussed below,
demonstrating that that our choice for the parameters are consistent
with the latest empirical results.
The thick black curves show the prediction 
for the Sheth-Tormen mass function for ellipsoidal collapse of DM haloes \citep{ShethTormen02},
whereas the thin grey curves show that
for the Press-Schechter mass function for spherical collapse \citep{PressSchechter74}.
The former is known to give better agreement with cosmological $N$-body simulations,
especially at the massive end of the mass function \citep[e.g.][]{Reed+07}.
Panel (b) shows the comoving number density of atomic-cooling haloes
as a function of $z$, i.e.
\beq
n_{\rm halo}(\Tvir>8000\,\K, z)=\int_{M(8000\,\K)}^\infty \frac{dn}{dM^\prime} ~dM^{\prime}.
\eeq
The line styles denote the assumed cosmological parameters
and the mass function in the same way as in panel (a).

Panel (c) shows $p(>v)$,
the probability that a random point in space
has a recombination value of the streaming velocity above $v$. This is simply the cumulative
distribution Maxwell-Boltzmann distribution function with an rms value of $30\,\kms$.
Finally, in panel (d) we present the comoving number density
of PAC haloes, i.e. those haloes with $\Tvir>8000\,\K$
and that have not yet formed stars
due to their lying in a region in space with a large streaming velocity:
\beq
n_{\rm halo}({\rm PAC})
=
\int_{M(8000\,\K)}^\infty \frac{dn}{dM^\prime}~p\left(>v_{8000} \right)~dM^{\prime},
\eeq
where $v_{8000}$ is the streaming velocity value at which $M_{\rm cool}$
exceeds the halo mass corresponding to $\Tvir=8000\,\K$.
As in Fig.\ref{fig:masses}, the cases with $\alpha=3.2$, $4.0$ and $4.7$ are shown in
red, green and blue.
For simplicity, we have only plotted cases for the Sheth-Tormen mass function.

\begin{table}
  \begin{tabular}{|l|  c c c c c | } \hline
    Parameters & $h$ & $\Omega_0$ & $\Omega_{\rm b}$ & $\sigma_8$& $n_{\rm s}$ \\ \hline 
    adopted & $0.7$ & $0.3$ & $0.047$ & $0.83$ & $0.96$ \\
    \textit{WMAP}9 & $0.700$ & $0.282$ & $0.0469$  & $0.827$& $0.980$\\
    \textit{Planck} & $0.678$ & $0.309$ & $0.0483$  & $0.829$ & $0.961$\\ \hline
  \end{tabular}
  \caption{$\Lambda$CDM cosmological parameters used to compute
  the halo abundances in Fig. 2.}
  \label{tab:cosmo}
\end{table}

Note that the comoving number densities presented in panels b)
and d) of Fig. \ref{fig:rarities} do not account for the fact
that BSMs suppress the number density of DM haloes
at high redshifts (by a factor  $\sim 2$ at $z\ga 30$; \citealt{TseliakHirata10,Naoz+12}).
However, the number density of massive BHs formed
is far more sensitive to other theoretical uncertainties,
such as the actual mass scale on which
PopIII formation occurs in the high-$v_{\rm BSM}$
regime (i.e. the effective value of the parameter $\alpha$)
and the efficacy of the collisional-dissociation scenario proposed by \cite{IO12}.

Fig. \ref{fig:rarities} allows one to place simple
order-of-magnitude upper limits.
From panel (a), the comoving number density of DM haloes
massive enough to plausibly host the most massive
SMBHs ($>10^9\Msol$) at $z\approx 6$ is
$\sim10^{-6}$--$10^{-5} \,\Mpc^{-3}$, if we suppose the host halo must
have a minimum mass of $\sim 10^{12}\,\Msol$.
Comparing panel (c) of this figure with Fig. \ref{fig:masses},
we can estimate the likelihood of any such halo having grown
in a region with sufficiently large streaming velocities to
be PAC as $\sim 10^{-3}$--$10^{-4}$.
Then, the comoving number density of haloes that could
host very massive SMBHs at $z=6$ and 
that had streaming velocities high enough to
have PAC progenitors may be as large as $\sim 10^{-9}$--$10^{-8}\,\Mpc^{-3}$,
i.e. large enough to account for the most massive SMBHs (most luminous quasars) at $z\ge 6$ \citep{Willott+10b}.

\subsection{Merger-tree simulations}
\label{subsec:trees}

We use ellipsoidal-collapse DM merger trees \citep{Zhang+08b}
convolved with the BSM velocity distribution
to estimate the number density of massive BHs formed in fast-streaming PAC haloes.
As stated above,
BSMs suppress the abundance of high-mass haloes, but
incorporating this effect in a merger tree code is a mathematically complex task.
Therefore, for practical reasons, we treat the DM mass function and
the BSM velocity fluctuations as being independent from each other.
As a result, our simulations slightly overestimate the halo abundance,
but this effect should be no more than a factor of $\sim 2$ at $z<30$.
The halo abundances also have numerical errors, typically less than a factor of 2
(see Fig. \ref{fig:halo} below).
As discussed above, these uncertainties are much smaller than
those associated with PopIII formation and the efficacy of direct collapse in PAC gas clouds.
We therefore do not believe that this simplification qualitatively affects our findings.

We simulate the merger histories of haloes with $z=6$ masses
$M_{\rm halo}\ge 10^8\,\Msol$.
Our sample includes 60 individual haloes with
$M_{\rm halo}>10^{12.9}\,\Msol$ at $z=6$,
equivalent to a comoving volume of $\approx 50 \,{\rm Gpc}^3$.
The merger tree sampling method and algorithm are described in more detail in
\citeauthor{TH09} (\citeyear{TH09}; section 2.6) and \cite{TLH13}, respectively;
we refer the reader to these works for details.
One key difference is that whereas in \cite{TLH13} a randomly generated
streaming velocity value was assigned to each merger tree, here
we convolve the halo sample with the velocity probability distribution function.
More explicitly, the halo mass function is computed
by counting the number of haloes in a given mass bin $M_{\rm lo}<M_{\rm halo}<M_{\rm hi}$
and dividing the sum by the effective comoving volume $V$ of the simulation sample and by the logarithmic
size of the bin:
$\phi_{\rm halo}(M_{\rm halo},z)=N(M_{\rm lo}<M_{\rm halo}<M_{\rm hi}, z)/V/\log_{10}(M_{\rm hi}/M_{\rm lo})$.
Note that the simulated value of $n_{\rm halo}$ depends both
on the fidelity of the merger tree algorithm in reproducing the theoretical
mass function (which we have discussed in the first paragraph of this subsection),
as well as on sample variance.
As long as the sampling error is small (i.e. $N\gg 1$ for the given mass bin),
given a sufficiently large volume the number density of PAC haloes is given by
\beq
\phi_{\rm halo}^{\rm (PAC)}(M_{\rm halo},z)=\sum  p_{\rm PAC}(z)/V/\log_{10}(M_{\rm hi}/M_{\rm lo}),
\label{eq:convolve}
\eeq
where the sum is performed over the mass bin
and $p_{\rm PAC}$ is the probability that a given halo resides in a region
of space where the streaming velocity has a value
such that the halo has had at least one PAC progenitor in its merger history.

In a given time step, a halo can become PAC if
(i) its virial temperature exceeds $8000\,\K$
and
(ii) the local streaming velocity is large enough (as defined by equation \ref{eq:Mcool})
so that it has not previously formed PopIII stars;
and
(iii) the velocity is small enough for gas infall and cooling to have occurred.
These conditions define a range or `window' of streaming velocity
values for which a given halo could have produced a massive BH;
i.e., if the velocity is too low, the halo will already have formed stars before
reaching $\Tvir=8000\,\K$,
and if it is too high, it will not have experienced gas infall and cooling (yet).
When two atomic-cooling haloes merge, their
velocity windows are also merged inclusively---e.g.
if one progenitor could have formed a massive BH in the streaming
velocity window $70\,\kms < v_{\rm BSM}^{\rm (rec)}< 80\,\kms$
and the other has a window $85\,\kms < v_{\rm BSM}^{\rm (rec)}< 90\,\kms$,
then the merged halo will have at least one PAC progenitor
if the local streaming velocity lies in either of these windows.
The probability $p_{\rm PAC}$ that the local $v_{\rm BSM}$ value lies within the combined set of velocity windows
is used to compute the abundance of PAC haloes
(equation \ref{eq:convolve}).

\begin{figure}
\epsfig{file=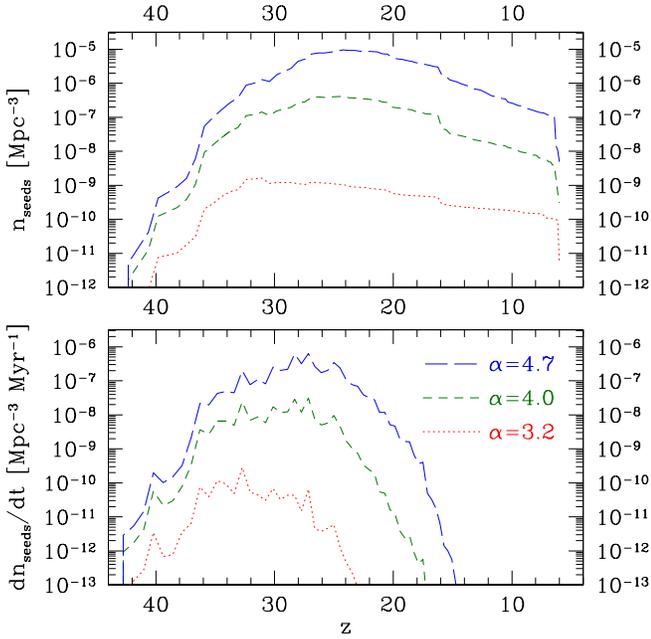,scale=0.44}
\vspace{-0.2in}
\caption{
Top panel:
estimated comoving number densities, as a function of $z$,
of DM haloes containing at least one massive BH
formed in PAC gas in rare, fast-streaming regions.
Bottom panel:
the rate at which haloes form massive BHs in this way.
See text for caveats.
(A color version of this figure is available online.)
}
\label{fig:seeds}
\end{figure}

To keep this analysis as model-independent as possible,
we make no \textit{a priori} assumptions of the BH accretion rate,
and only keep track of the number densities and mass functions
of haloes that host at least one PAC progenitor.
We also do not account for possible ejections of massive BHs
via the gravitational recoil effect of BH mergers
\citep{Peres62, Bekenstein73,Favata+04,Haiman04,YooME04,
Baker+06,SchnittBuon07,BlechaLoeb08, Guedes+08, TH09}.
The latter assumption is likely justified; the host DM haloes of
interest have deeper potentials than those of PopIII stars
and inhibit ejections.
Further, even if a large fraction of BH pairings result in
ejections, this should not affect the BH occupation fraction
of massive haloes, as a typical $M>10^{12}\,\Msol$ halo at $z=6$
in a fast-streaming patch of space has
$N_{\rm prog}\sim 10^2$--$10^3$ PAC progenitors that could have formed a massive BH.
This can be seen from the top panel of Fig.\ref{fig:seeds},
where the number density of haloes with PAC progenitors
decreases by a factor $10^2$--$10^3$ for each $\alpha$ value considered.

An important related point is that massive $z\ltsim 10$ haloes
that assembled in a fast-streaming region is likely to host
a massive BH even if the fraction $f_{\rm DC}$ of PAC haloes---i.e. \textit{potential} direct-collapse sites---that
\textit{actually} result in direct collapse is small.
Again, the massive $z\ltsim 10$ haloes
have consumed $N_{\rm prog}\gg 1$ PAC haloes via mergers.
This means that as long as $f_{\rm DC}\ga 1/N_{\rm prog}$,
the number density of massive BHs at $z\ltsim 10$
is roughly independent of $f_{\rm DC}$.
That low BH occupation fractions can result in occupation fractions of
unity at later times was shown by \cite{Menou+01} and \cite{TH09}.

Fig. \ref{fig:seeds} shows the total number density of DM haloes
that could host massive BHs formed in
PAC haloes with large streaming velocities.
The top panel shows the total comoving number density of host
haloes as a function of redshift; note that at late times ($z<20$),
the number density goes down through hierarchical merging of haloes.
The bottom panel shows the global seed formation rate.
As with the previous figures, each curve assumes a different value of the parameter $\alpha$
that determines the threshold mass for PopIII formation:
red dotted for $\alpha=3.2$, green short-dashed for $\alpha=4.0$
and blue long-dashed for $\alpha=4.7$.
The formation and merger rates of massive BHs made in the scenario
considered here would be very low, with peak all-sky formation rates
$\ltsim 10^{-3}\,{\rm yr}^{-1}$ per unit redshift,
or no more than $\sim 10^{-2}\,{\rm yr}^{-1}$ integrated across all redshifts.
Even if these $z>20$ massive BHs formed in binary or multiple systems \citep[e.g.][]{BrommLoeb03},
their mergers are unlikely to be observed by
a gravitational-wave observatory such as \textit{eLISA} \citep{eLISApaper} during its mission lifetime.

Fig. \ref{fig:halo} shows the mass function of all DM haloes
(solid black curves and histograms) and of those containing massive BHs
formed in fast-streaming regions (colored
 curves), at redshifts $z=32$, 24, 17, 11, 7.1 and 6.
The \textit{James Webb Space Telescope}\footnote{
\url{http://www.jwst.nasa.gov/}} (\textit{JWST}) is expected to
be capable of detecting early quasars at $z\approx 11$;
however, these massive BHs will have a sky density of $\ltsim 0.1 \,{\rm dex}^{-1} \,{\rm deg}^{-2}$
per unit redshift (see, e.g., figure 4 in \citealt{TLH13} for conversion of comoving $n$ to sky density at $z=11$)
and thus are unlikely to be discovered by \textit{JWST},
whose field of view will be $\sim 10^{-3}\,{\rm deg}^2$.
The solid black curve shows the expected mass function
in the Sheth-Tormen formalism, whereas the solid black histograms show
the number densities produced by the merger trees.
The differently colored histograms show the mass function
of haloes containing at least one massive BH seed,
with different colors and line styles showing the different assumed values
for $\alpha$ as in all of the previous figures.
As anticipated in \S\ref{subsec:semianalytic}, the number density
of $z\approx 6$--$7$ haloes with $M>10^{12}\,\Msol$ that contain massive BHs
formed via large streaming velocities could be as large as $\ga 10^{-9}\,\Mpc^{-3}$,
i.e. large enough to account for the most massive quasar SMBHs observed at $z>6$.

\begin{figure}
\epsfig{file=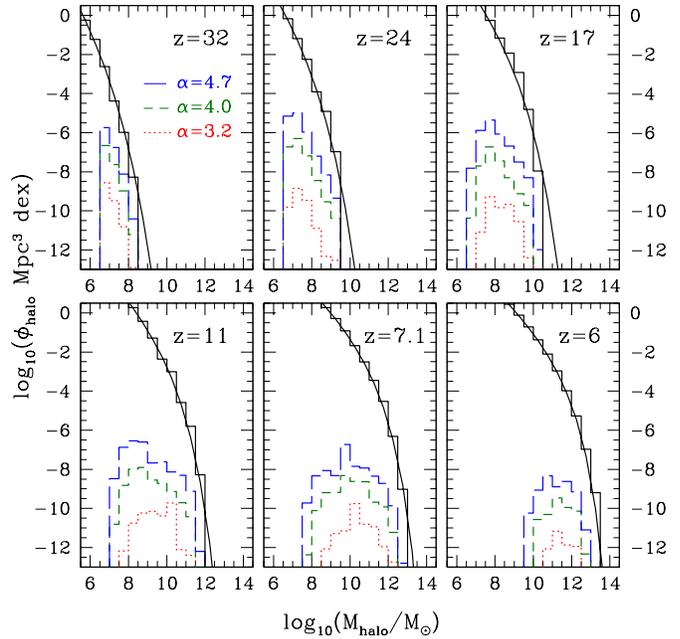,scale=0.44}
\vspace{-0.2in}
\caption{
The comoving number densities
of PAC haloes formed in fast-streaming regions of the Universe,
assuming halo mass thresholds for metal contamination
in equation \eqref{eq:Mcool} for values of $\alpha=3.2$
(red dotted lines),   $\alpha=4.0$ (greed short-dashed lines)
and $\alpha=4.7$ (blue long-dashed lines).
The solid black curves and histograms show, respectively, the
mass functions of DM haloes expected by the Sheth-Tormen
formalism and those produced by the merger trees.
(A color version of this figure is available online.)
}
\label{fig:halo}
\end{figure} 

\section{Discussion and Conclusions}
\label{sec:disc}

We have proposed in this paper a path to massive BH formation via direct collapse
of pristine, atomic-cooling (PAC) gas at $z>20$.
The mechanism has two sequential prerequisites,
both of which have been discussed in the recent literature.
First, large baryonic streaming velocities \citep{TseliakHirata10} must be able
to delay PopIII formation \citep{Stacy+11, Greif+11a, Fialkov+12, Naoz+13} until the halo gravitational potential
is deep enough to host atomic-cooling gas, i.e. until $\Tvir\ga 8000\,\K$.
Second, the gas that finally accumulates inside this halo must then undergo
direct collapse, e.g. by forming a central cloud that is hot and dense enough ($T\ga 10^4\,\K$, $n\ga10^3\,\cm^{-3}$)
to collisionally dissociate $\Hmol$ and collapse via atomic cooling.
In essence, the first condition facilitates the conditions for
UV-free direct-collapse by providing a natural mechanism for forestalling PopIII formation
and enrichment by metals and dust.

The feasibility of the first condition occurring in nature has been
demonstrated, at least qualitatively, by recent three-dimensional simulations.
Essentially, PopIII formation proceeds very rapidly via runaway $\Hmol$ and ${\rm HD}$
cooling when the central gas reaches certain density thresholds inside a halo
with $\Tvir\ga 1000\,\K$, and streaming velocities delay this transition.
The key question, however, is just how large the streaming velocity must
be to delay PopIII formation until the halo can host atomic-cooling gas,
and whether such velocities occur frequently enough
to explain the most massive of the $z>6$ quasar SMBHs.
If one extrapolates the increase in the minimum PopIII-forming halo mass
found by \cite{Stacy+11} and \cite{Greif+11a} to large streaming velocities, as \cite{Fialkov+12} do,
then the requisite velocity to stall PopIII formation until haloes become atomic-cooling
is about two to three times the rms value ($\ga 60 \,\kms$; Fig. \ref{fig:masses}).
\cite{Naoz+13} show that at such streaming velocities, gas fractions
in $\Tvir\sim 8000 \,\K$ haloes are indeed suppressed at $z\ga 20$.
These two sets of findings show that at least in principle, the first condition could be fulfilled in nature.
However, there are significant uncertainties that have not been
addressed by high-resolution simulations.
As pointed out in \S\ref{sec:masses}, dynamical and compressional heating
would be enhanced in the more massive, gas-poor haloes affected by large streaming velocities.
Furthermore, the role of turbulence---which streaming velocities enhance---in promoting or suppressing
PopIII formation has not been explored by simulations for the rare set of conditions discussed here.

A more general question is whether
a PAC cloud can lead to direct collapse at all.
While several studies have demonstrated the plausibility
that direct collapse can occur in the absence of an external UV background,
i.e. not via photodissociation but via collisional dissociation,
detailed tests in numerical simulations are yet to come.
Such conditions may lead to a `supermassive' star
that explodes instead of collapsing into a massive BH (see \citealt{Johnson+13} and references therein).
If UV-free direct-collapse occurs in nature at all, the streaming motions
can facilitate it (i) by minimizing metal enrichment or even preventing it entirely,
and (ii) through enhanced supersonic turbulence, promoting the creation of cold accretion filaments
and increasing their kinetic energies.
As argued above, 
the mechanism proposed here can result in
a comoving number density $\sim 10^{-9}$--$\sim 10^{-8}\,\Mpc^{-3}$
of $z\approx6$--$7$ haloes containing a massive BH
\textit{even if only a small fraction ($\ltsim 1$ per cent)
of PAC haloes actually result in direct collapse}.
The small fraction could be those that have low angular momentum
or those that are clustered near strong UV sources (see refs. in \S\ref{sec:intro}).

Detailed hydrodynamical simulations are required
to resolve both of these substantial theoretical uncertainties.
High-resolution simulations of $\Tvir\ltsim 10^3\,\K$ DM haloes growing
in large density peaks and amidst large streaming motions
are needed to verify whether such conditions can forestall or minimize
PopIII formation and metal enrichment until the halo potential is deep enough
to support atomic-cooling gas and possibly shocking filaments.
The simulations must then follow the gas in these haloes
to confirm whether direct collapse can occur,
as well as determine the degree to which collapse is sensitive to the
metal content of the halo gas and to the magnitude of the streaming motions.

Of particular interest is the role of turbulence,
which has been shown to be important in the formation of baryonic structures
in both $\Tvir\sim 1000\,\K$ and $\Tvir\sim 10^4\,\K$ haloes
\citep{Greif+08, Greif+11b}.
\cite{Fernandez+13} found, in simulations
of atomic-cooling haloes with pristine gas but without streaming velocities,
that while gas in these haloes indeed do not cool efficiently,
they also do not form the cool supersonic filaments envisioned by \cite{IO12}.
It remains to be seen if such filaments can form with large streaming
velocities, i.e. if the gas is more turbulent; similarly, the effects of
supersonic turbulence on other proposed direct-collapse scenarios is uncertain.

If large baryonic streaming velocities can indeed stall PopIII formation
until the earliest massive haloes reach $\Tvir\sim 8000 \,\K$
(if $\alpha\ga4$, as \citealt{Fialkov+12}  find for \citealt{Greif+11a}),
then these haloes will have gas that is PAC and exceptionally turbulent,
making them promising potential cradles of direct collapse
or exceptionally massive stars.
These sites would preferentially emerge in a redshift range $20\ltsim z\ltsim 40$
(Fig. \ref{fig:rarities} panel d, Fig. \ref{fig:seeds} bottom panel);
at higher redshifts, haloes with $\Tvir\ga 8000\,\K$ are too rare,
whereas at lower redshifts streaming velocities are too low to delay PopIII formation.
This is much earlier than the direct-collapse scenarios
that rely on the emergence of UV sources at $z<16$,
and would produce rare $\sim 10^5\,\Msol$ BHs alongside the very first galaxies.
The minimum time-averaged accretion rate required for such objects
to grow into the observed $z>6$ SMBHs is
\begin{align}
f_{\rm Edd}\ga
&\left[0.499+0.048\ln\left(\frac{M_{\rm SMBH}}{3\times10^9\,\Msol} \frac{10^5\,\Msol}{M_{\rm seed}}\frac{1}{X_{\rm merge}}\right)
\right] \nonumber\\
&\qquad \times\left(\frac{\eta}{0.07}\right)\left(\frac{t_{\rm avail}}{650\,\Myr}\right)^{-1}.
\label{eq:fEddBSM}
\end{align}
Comparing equation \eqref{eq:fEddBSM} above
to equations \eqref{eq:fEddPopIII} and \eqref{eq:fEddUV},
we see that these massive seed BHs
offer as much of an advantage over direct-collapse seeds formed via
large UV backgrounds as the latter do over PopIII seeds.

Several ways to observationally distinguish PopIII and direct-collapse models
have been discussed in the literature
(see \citealt{Volonteri10}, \citealt{Haiman13}).
There is a degeneracy between most direct-collapse models and PopIII scenarios,
in that both require (if the mean accretion rate does not far exceed the Eddington value)
that $\sim 10^5\,\Msol$ BHs are in place before $z\approx 10$ 
(see, e.g., section 4.2.1 in \citealt{TPH12}).
Directly breaking this degeneracy---i.e. probing nuclear BHs at $z>10$---will be extremely challenging,
even with upcoming missions such as \textit{JWST} and \textit{Athena+}.
It is also possible that SMBH progenitors are so rare \citep[e.g.][]{TH09}
that they are unlikely to be observed in gravitational waves or through explosions
of `supermassive' stars.
That being said, the scenario presented here can be corroborated if ever a massive
($>10^4\,\Msol$) BH---or an associated signature, such as the explosion of a supermassive progenitor star
or gravitational-wave signature---is discovered at $z>16$ (where UV-aided direct collapse is unlikely).

Direct-collapse BHs are expected to have much larger masses with respect to
their host halo mass than typical nuclear BHs (\citealt{BrommLoeb03}, \citealt{Agarwal+13}).
They could be the progenitors of present-day SMBHs with
unusually large masses compared to their galaxies,
as in NGC 1277 \citep{vdBosch+12, ShieldsBonning12}.

While a large baryonic streaming velocity cannot be the only pathway to SMBH
formation (since SMBHs are present in virtually all galaxies, not just
in rare patches of the Universe that had large BSMs),
it can explain how a small number of SMBHs were able
to grow to be exceptionally massive before  $z\approx 6\,$--$\,7$.

\section*{Acknowledgements}
We thank Zolt\'an Haiman for detailed comments on the manuscript,
and the anonymous reviewer for helpful suggestions.
TLT is grateful to Greg Bryan, Mark Dijkstra, Thomas Greif,
Zolt\'an Haiman, Jeremiah Ostriker, Andreas Pawlik and Rashid Sunyaev for insightful discussions.


\end{document}